# ACOUSTIC ECHO SUPPRESSION USING A LEARNING-BASED MULTI-FRAME MINIMUM VARIANCE DISTORTIONLESS RESPONSE (MFMVDR) FILTER

*Yuefeng Tsai[1], Yicheng Hsu[1], and Mingsian R. Bai[1,2]*

[1]Department of Power Mechanical Engineering, National Tsing Hua University, Taiwan
[2]Department of Electrical Engineering, National Tsing Hua University, Taiwan

## ABSTRACT

Distortion resulting from acoustic echo suppression (AES) is a common issue in full-duplex communication. To address the distortion problem, a multi-frame minimum variance distortionless response (MFMVDR) filtering technique is proposed. The MFMVDR filter with parameter estimation which was used in speech enhancement problems is extended in this study from a deep learning perspective. To alleviate numerical instability of the MFMVDR filter, we propose to directly estimate the inverse of the correlation matrix. The AES system is advantageous in that no double-talk detection is required. The negative scale-invariant signal-to-distortion ratio is employed as the loss function in training the network at the output of the MFMVDR filter. Simulation results have demonstrated the efficacy of the proposed learning-based AES system in double-talk, background noise, and nonlinear distortion conditions.

## 1. INTRODUCTION

Excessive acoustic echo could result in severe degradation of full-duplex communication quality. Due to the coupling between the loudspeaker and the microphone at the near-end, the far-end user may hear the near-end speech corrupted by the echo. To tackle this issue, two classes of approaches, including the acoustic echo cancellation (AEC) [1, 2] or acoustic echo suppression (AES) [3, 4] algorithms, have been proposed. The common goal of these algorithms is to eliminate the echo component in the mixture signal, while preserving the near-end signal with minimal distortion.

AEC algorithms utilize linear adaptive filters to estimate the acoustic echo path and subtract the estimated echo signal from the microphone signal. These methods require a double-talk detector [5, 6] to ensure stable filter adaptation when users at both ends speak concurrently. For small loudspeakers used in many consumer electronic devices, nonlinear distortion due to overdriving transducers may further degrade the cancellation performance. To remedy, post-filtering [7] and residual echo suppression (RES) [8, 9, 10] methods were suggested in the literature.

As opposed to the AEC approaches, AES systems are based on spectral subtraction principle and hence do not require double-talk detection [11, 12]. In these methods, echo components need to be estimated by taking advantage of the property that the echo signal and the near-end signal are uncorrelated. The undesired echo signal is attenuated accordingly, while preserving the near-end speech during the double-talk period. However, for low Signal-to-Echo Ratio (SER) scenarios, the suppression performance of AES could deteriorate significantly.

Recently, neural networks have shown great promise in acoustic echo suppression, where nonlinearity is generally better handled. Wang et al. [13] adopted the weighted Recursive Least Square (wRLS) filter for the linear part of the echo and a neural network for the nonlinear part. Zhang and Wang [14] formulated AEC as a source separation problem and applied a bidirectional long-short term memory (BLSTM) to predict an ideal ratio mask to extract the near-end speech. After then, various networks were suggested to address AEC problems [15, 16].

However, how to control speech distortion properly remains to be a challenge. To alleviate the speech distortion resulting from AES, a novel technique based on multi-frame minimum variance distortionless response (MFMVDR) [17] filtering is proposed. MFMVDR exploits the interframe speech correlation and reformulates the single-channel problem with the notion akin to array beamforming to prevent the target speech from distortion. In the MFMVDR filter, the noise correlation matrix (NCM) and the speech interframe correlation (IFC) are computed. The AES performance relies on reliable estimation of NCM and IFC, in particular the latter. Recently, some learning-based techniques were suggested to estimate these two parameters [18, 19].

The contributions of this study are summarized as follows. We extend the MFMVDR filter to acoustic echo suppression, with the aid of a parameter estimation network. We develop the MFMVDR filter within the deep learning framework as suggested in [19]. Instead of predicting the correlation matrix and the *a prior* SNR, the proposed network directly predicts the correlation matrix inverse and the target speech IFC. The negative scale-invariant signal-to-distortion ratio (SI-SDR) is employed as the loss function in training the network at the output of the MFMVDR filter [20].

The remainder of this paper is organized as follows. We first define the signal model for the MFMVDR filter in Section 2. The learning-based MFMVDR model is presented in Section 3. The results are discussed in Section 4. Conclusions are given in Section 5.

## 2. SIGNAL MODEL

An acoustic echo suppression system is depicted in Figure 1, where the noisy microphone signal $y(n)$ is the mixture of near-end signal $s(n)$, acoustic echo $d(n)$, and background noise $v(n)$. We assume the near-end signal, the acoustic echo, and the background noise are mutually uncorrelated and zero-mean signals. In the Short Time

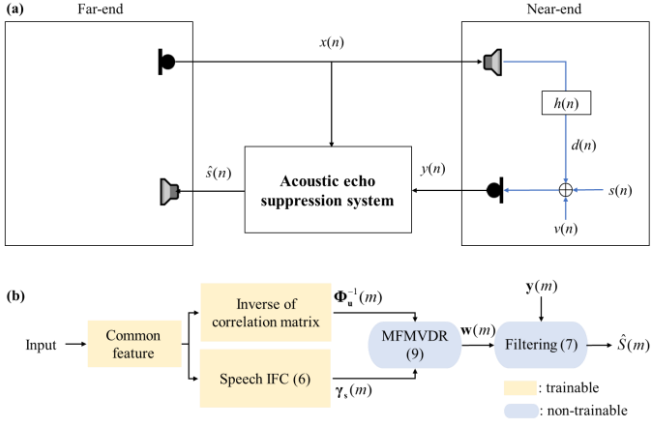

**Fig. 1.** Acoustic echo suppression (AES). (a) The AES problem and (b) the block diagram of the proposed AES system.

Fourier Transform (STFT)-domain, the noisy microphone signal is given by

$$Y(k,m) = S(k,m) + D(k,m) + V(k,m), \quad (2)$$

where $Y(k,m)$, $S(k,m)$, $D(k,m)$, and $V(k,m)$ denote the STFTs of respective signals at the $k$-th frequency bin and the $m$-th time frame.

The multi-frame processing exploits the correlation of speech signal across time frames. First, we define the noisy signal vector of the current noisy frame as

$$\begin{aligned}\mathbf{y}(k,m) &\triangleq [Y(k,m) \cdots Y(k,m-L+1)]^T \\ &= \mathbf{s}(k,m) + \mathbf{d}(k,m) + \mathbf{v}(k,m),\end{aligned} \quad (3)$$

where $L$ represents the number of frames and the superscript $T$ denotes the transpose operator. Vectors, $\mathbf{s}(k,m)$, $\mathbf{d}(k,m)$, and $\mathbf{v}(k,m)$ are defined similarly. The $L \times L$ noisy correlation matrix can be expressed as

$$\begin{aligned}\mathbf{\Phi}_\mathbf{y}(k,m) &\triangleq E\{\mathbf{y}(k,m)\mathbf{y}^H(k,m)\} \\ &= \mathbf{\Phi}_\mathbf{s}(k,m) + \mathbf{\Phi}_\mathbf{d}(k,m) + \mathbf{\Phi}_\mathbf{v}(k,m) \\ &\triangleq \mathbf{\Phi}_\mathbf{s}(k,m) + \mathbf{\Phi}_\mathbf{u}(k,m),\end{aligned} \quad (4)$$

where $E\{\bullet\}$ represents mathematical expectation and the superscript $H$ denotes the conjugate-transpose operator. Then, $\mathbf{\Phi}_\mathbf{s}(k,m)$, $\mathbf{\Phi}_\mathbf{d}(k,m)$, and $\mathbf{\Phi}_\mathbf{v}(k,m)$ are the correlation matrices associated with $\mathbf{s}(k,m)$, $\mathbf{d}(k,m)$, and $\mathbf{v}(k,m)$, respectively. The correlation matrix of the undesired signal, $\mathbf{\Phi}_\mathbf{u}(k,m) \triangleq \mathbf{\Phi}_\mathbf{d}(k,m) + \mathbf{\Phi}_\mathbf{v}(k,m)$. The near-end speech signal vector $\mathbf{s}(n)$ can be separated into two orthogonal components corresponding orthogonal components corresponding to the desired signal and the undesired signal. That is,

$$\mathbf{s}(k,m) = \boldsymbol{\gamma}_S(k,m)S(k,m) + \mathbf{s}'(k,m), \quad (5)$$

where $\boldsymbol{\gamma}_S(k,m)$ is the IFC vector and $\mathbf{s}'(k,m)$ is the interference vector with respect to $S(k,m)$. This vector indicates the correlation between current time frame signal $S(k,m)$ and the previous time frame signals, $S(k, m-l)$, $l = 0,1,\ldots,L-1$, and can be viewed as the steering vector in an array beamformer. Specifically,

$$\boldsymbol{\gamma}_S(k,m) = \frac{E\{\mathbf{s}(k,m)S^*(k,m)\}}{E\{|S(k,m)|^2\}} = \frac{\mathbf{\Phi}_\mathbf{s}(k,m)\mathbf{e}}{\mathbf{e}^T\mathbf{\Phi}_\mathbf{s}(k,m)\mathbf{e}}, \quad (6)$$

where superscript * represents the complex conjugate operator and $\mathbf{e} = [1,0,\ldots,0]^T$ is an $L$-dimensional one-hot vector.

## 3. THE LEARNING-BASED MFMVDR

### 3.1. MFMVDR

The MFMVDR filter is employed to recover the clean signal by applying the weight vector $\mathbf{w}$ to the noisy signal:

$$\hat{S}(k,m) = \mathbf{w}^H(k,m)\mathbf{y}(k,m). \quad (7)$$

With the distortionless constraint, the MFMVDR filter weights can be obtained by solving the following constrained optimization problem:

$$\begin{aligned}\min_{\mathbf{w} \in \mathbb{C}^L} & \quad \mathbf{w}^H(k,m)\mathbf{\Phi}_\mathbf{u}(k,m)\mathbf{w}(k,m) \\ \text{s.t.} & \quad \mathbf{w}^H(k,m)\boldsymbol{\gamma}_S(k,m) = 1\end{aligned} \quad (8)$$

which yields the optimal MFMVDR filter vector

$$\mathbf{w}_{MFMVDR}(k,m) = \frac{\mathbf{\Phi}_\mathbf{u}^{-1}(k,m)\boldsymbol{\gamma}_S(k,m)}{\boldsymbol{\gamma}_S^H(k,m)\mathbf{\Phi}_\mathbf{u}^{-1}(k,m)\boldsymbol{\gamma}_S(k,m)}. \quad (9)$$

### 3.2. DNN structure

In this section, we use a deep neural network (DNN) to estimate the parameters required in the MFMVDR filter design. To bypass the computation of the matrix inverse and the correlation vector in (9) and (6), the proposed network directly estimate the undesired signal correlation matrix inverse ($\mathbf{\Phi}_\mathbf{u}^{-1}$) and the target speech IFC ($\boldsymbol{\gamma}_S$), based on the far-end signal and the noisy signal.

Figure 2 illustrates the parameter estimation network for the MFMVDR system. The network consists of a shared section and a task-specific section. The shared section receives the real and imaginary parts of the STFTs of the far-end signal and the noisy signal and extracts common features for the correlation matrix inverse and IFC. The shared section is comprised of a four-layer temporal convolutional network (TCN) [22] with hidden units of 256 and a kernel size of 3, which is very effective in temporal and spectral modeling. The task-specific section has two parallel outputs: the correlation matrix inverse estimate and the IFC estimate. A three-layer TCN module with hidden units of 256 and a kernel size of 3. Furthermore, the TCN module is cascaded with a complex gated recurrent unit (cGRU) with 96 hidden units. The cGRU structure is inspired by complex long-short term memory (cLSTM) in [23]. Complex multiplication is carried out in terms of the real part ($X_r$) and the imaginary part ($X_i$) of the complex input. The output of cGRU output $F_{out}$ can be written as

$$F_{out} = (F_{rr} - F_{ii}) + j(F_{ri} + F_{ir}) \quad (10)$$

where

$$F_{rr} = \text{GRU}_r(X_r); \quad F_{ir} = \text{GRU}_r(X_i) \quad (11)$$

$$F_{ri} = \text{GRU}_i(X_r); \quad F_{ii} = \text{GRU}_i(X_i) \quad (12)$$

with $\text{GRU}_r$ and $\text{GRU}_i$ denoting two GRUs associated with the real and imaginary parts.

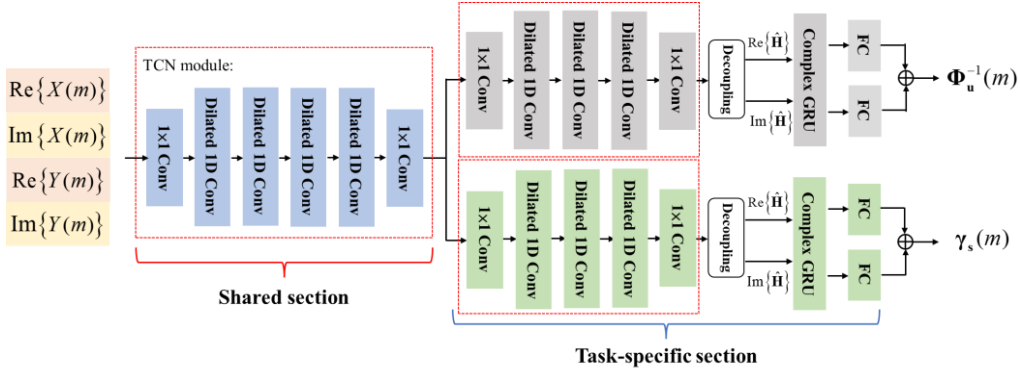

**Fig. 2.** Parameter estimation network for the MFMVDR system and the TCN module marked with red-dotted boxes.

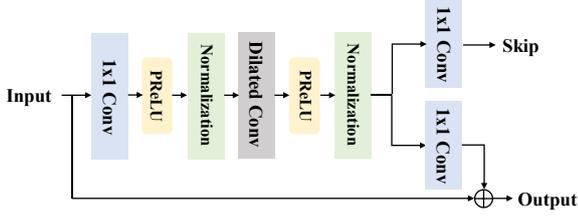

**Fig. 3.** Dilated 1D convolution block

Important features are first extracted by the corresponding TCN module. Next, we use the technique in [20] to separate the output into two $L^2$-dimensional vectors, including the real part and the imaginary part of the correlation matrix. Lastly, the cGRU module processes the temporal characteristics and feeds them to a fully connected (FC) layer to produce the output parameters, the correlation matrix inverse and IFC. The network is trained to minimize the SI-SDR loss to be defined in Sec. 4.1 at the MFMVDR output.

## 4. SIMULATIONS

### 4.1. Data augmentation and training settings

To validate the proposed method, two public datasets, Librivox [25] and openSLR26 [26] from the INTERSPEECH 2021 Deep Noise Suppression (DNS) datasets [27] are used. We randomly select 45000 pairs of utterances sampled at 16kHz from the Librivox dataset to serve as the near-end and far-end source signals. In addition, 45000 room impulse responses (RIRs) chosen from openSLR26 is used as the acoustic echo. Each far-end signal is 4 s in length. The near-end signal is a 2 s clip and zero-padded to the same size. Then, the far-end signal is convolved with one of the RIRs to form the acoustic echo signal, limited to 4 s in length. The microphone signal is the mixture of the acoustic echo and the near-end signal, with signal-to-echo (SER) and signal-to-noise ratio (SNR) randomly selected in [-20, 10] dB and [10, 40] dB, respectively. Using a 20% validation split, this result in 40000 and 5000 mixture signals for training and validation. We also apply the method suggested in [14] to simulate the distortions caused by loudspeaker nonlinearity.

The signal clips in length of 8ms and 4 ms are adopted in training. A 128-point STFT is applied to each time frame. A Hanning window was employed both as the analysis and synthesis window. The parameter estimation model of MFMVDR was trained with filter lengths, $L = 3, 5,$ and 7, to validate the proposed method.

The number of training epochs for the proposed network is set to 50. Adam optimizer [24] with an initial learning rate of 0.0003 is used. The learning rate is decreased by 1.5% every epoch. The gradient norms are clipped to 5 if exceeded. The negative scale-invariant signal-to-distortion ratio (SI-SDR) is adopted as the loss function.

$$\text{SI-SDR} = 10\log_{10}\left(\frac{|\alpha\mathbf{s}|^2}{|\alpha\mathbf{s} - \hat{\mathbf{s}}|^2}\right), \quad \alpha = \frac{\hat{\mathbf{s}}^T\mathbf{s}}{\|\mathbf{s}\|^2}, \quad (13)$$

where $\mathbf{s}$ and $\hat{\mathbf{s}}$ represent the clean and estimated time-domain signal vector.

**Table 1.** Averaged PESQ, STOI, and SI-SDR in double talk, sensor noise, and nonlinear distortion conditions, with SER= {-10, -5, 0} dB and SNR = 30 dB. MFMVDR-AES-*L* denotes a filter of length *L*.

| SER(dB) | | -10dB | | | -5dB | | | 0dB | | |
|---|---|---|---|---|---|---|---|---|---|---|
| Method | #Params | PESQ | STOI | SI-SDR | PESQ | STOI | SI-SDR | PESQ | STOI | SI-SDR |
| None | | 1.445 | 0.582 | -17.715 | 1.682 | 0.704 | -12.752 | 1.989 | 0.809 | -7.722 |
| Baseline | 2.9M | 2.115 | 0.783 | 9.823 | 2.592 | 0.865 | 13.140 | 3.058 | 0.920 | 16.416 |
| MFMVDR-AES-3 | 3.2M | 2.159 | 0.812 | 11.743 | 2.609 | 0.886 | 15.238 | 3.023 | 0.931 | 18.444 |
| MFMVDR-AES-5 | 3.5M | **2.198** | **0.817** | **11.925** | **2.671** | **0.891** | **15.419** | **3.083** | **0.934** | **18.592** |
| MFMVDR-AES-7 | 3.9M | 2.118 | 0.801 | 11.584 | 2.566 | 0.879 | 15.093 | 2.969 | 0.926 | 18.203 |

**Table 2.** Averaged ERLE in single-talk, sensor noise, and nonlinear distortion conditions, with SER= {-10, -5, 0} dB and SNR = 30 dB.

| | SER(dB) | -10dB | -5dB | 0dB |
|---|---|---|---|---|
| ERLE | Baseline | **54.435** | **51.851** | **49.049** |
| | MFMVDR-AES-3 | 36.077 | 36.435 | 35.431 |
| | MFMVDR-AES-5 | 34.747 | 35.102 | 33.877 |
| | MFMVDR-AES-7 | 33.289 | 34.306 | 33.159 |

### 4.2. Baseline model

In this work, we use a baseline model provided by the 2021 AEC-Challenge [28]. A recurrent neural network with GRU employs the concatenated log power spectrum of the noisy signal and the far-end signal as the input, to produce a spectral suppression mask. The required STFTs are computed, based on 20 ms frames with a stride of 10 ms and 320-point discrete Fourier transform. The model is comprised of two GRU layers with 512 hidden units, followed by a FC layer with a sigmoid activation function. The predicted mask is point-wise multiplied with the magnitude spectrum of the noisy signal to suppress the far-end signal. As a final step, the time-domain signal is recovered via the inverse STFT of the estimated magnitude spectrum multiplied by the noisy phase term. The baseline model is trained by the same setup as the proposed network.

### 4.3. Results and discussions

Four performance metrics are adopted to evaluate the AES systems: Echo Return Loss Enhancement (ERLE) [1], perceptual evaluation of speech quality (PESQ) [29], short-time objective intelligibility (STOI) [30], and SI-SDR. PESQ, STOI, and SI-SDR are evaluated when double-talks are present, whereas ERLE is evaluated for single-talk scenarios.

$$\text{ERLE} = 10\log_{10}\left(\frac{\sum_n y^2(n)}{\sum_n \hat{s}^2(n)}\right). \quad (14)$$

All models are evaluated with 2000 unseen mixture signals in SER = {-10, -5, 0} dB, SNR = 30 dB, and nonlinear distortion conditions. The results summarized in Table 1 suggest that all network models yield significant enhancement over the unprocessed signal in several performance measures. The MFMVDR-AES model with filter length $L = 5$ performs the best in terms of all objective metrics under all conditions. The PESQ scores of MFMVDR-AES with filter length $L = 3$ and 7 are slightly lower than the baseline model in the condition with SER = 0 dB, but still considerably outperform the baseline model in STOI and SI-SDR scores.

Averaged ERLE scores obtained using the proposed and the baseline networks are summarized in Table 2. The baseline model exhibits impressive echo suppression capability at the cost of the other speech quality measures, as indicated in Table 1. The MFMVDR-AES model preserves the near-end speech quality, while maintaining perceptually acceptable ERLE above 30 dB. It can be observed from the results that the performance of the proposed MFMVDR-AES models is dependent on the filter length. When only three frames are considered, the PESQ attained by MFMVDR-AES is slightly lower than the baseline model, whereas STOI and SI-SDR attained by MFMVDR-AES model are higher than those attained by the baseline model. The distortionless constraint in the MFMVDR filter design proves effective in reducing the distortion of the near-end signal incurred by the AES system. Increasing the filter length to five in the MFMVDR-AES model leads to the best performance in both speech quality and intelligibility. However, increasing further the filter length to seven in the MFMVDR-AES model, all objective metrics drop significantly. All objective indices of the MFMVDR-AES model with filter length $L = 7$ are slightly lower than $L = 3$, which means the correlation relationship is weak as considering too many frames. Therefore, the best performance of MFMVDR-AES may be unable to be achieved if excessive or insufficient number of time frames are considered.

### 5. CONCLUSIONS

This paper has presented a novel learning-based AES system in light of a learning-based MFMVDR that exhibits superior performance over a baseline echo suppression approach. By exploiting the inter-frame correlation of the single-channel microphone signal, we are able to achieve excellent suppression performance with minimal distortion and artifact. This is mainly attributed to the fact that the DNN can yield reliable estimates of parameters required by the MFMVDR filter in the face of various adverse conditions. The overall system demonstrates the great synergy of state-of-the-art signal processing and emerging deep learning approaches.

### 7. ACKNOWLEDGMENT

This work was supported by the Ministry of Science and Technology (MOST), Taiwan, under the project number 110-2221-E-007-027-MY3.